\documentclass[conference]{IEEEtran}

\usepackage{algpseudocode}
\usepackage{amsmath}
\usepackage{amssymb}
\usepackage{booktabs}
\usepackage{bm}
\usepackage[nosort]{cite}
\usepackage[utf8]{inputenc}
\usepackage{csquotes}
\usepackage{graphicx}
\usepackage[final]{microtype}
\usepackage[caption=false,
            font=footnotesize,
           ]{subfig}
\usepackage{url}

\renewcommand\vec{\mathbf}
\newcommand{\norm}[1]{\left\lVert#1\right\rVert}

\graphicspath{{figures/}}

\title{Capsule Routing for Sound Event Detection}
\author{\IEEEauthorblockN{Turab Iqbal, Yong Xu, Qiuqiang Kong and Wenwu Wang}
        \IEEEauthorblockA{Centre for Vision, Speech and Signal Processing, University of Surrey\\
                          Email: \{t.iqbal, yong.xu, q.kong, w.wang\}@surrey.ac.uk}}

\begin{document}
  \bstctlcite{IEEEexample:BSTcontrol}

  \maketitle

  \begin{abstract}
    The detection of acoustic scenes is a challenging problem in which
    environmental sound events must be detected from a given audio signal. This
    includes classifying the events as well as estimating their onset and
    offset times. We approach this problem with a neural network architecture
    that uses the recently-proposed capsule routing mechanism. A capsule is a
    group of activation units representing a set of properties for an entity of
    interest, and the purpose of routing is to identify part-whole
    relationships between capsules. That is, a capsule in one layer is assumed
    to belong to a capsule in the layer above in terms of the entity being
    represented. Using capsule routing, we wish to train a network that can
    learn global coherence implicitly, thereby improving generalization
    performance. Our proposed method is evaluated on Task 4 of the DCASE 2017
    challenge. Results show that classification performance is
    state-of-the-art, achieving an F-score of 58.6\%. In addition, overfitting
    is reduced considerably compared to other architectures.
  \end{abstract}

  \section{Introduction}
  Sound event detection (SED) is the task of classifying and localizing sound
  events in audio such that each detected event is assigned a class label as
  well as onset and offset times. Recently, the problem has received
  significant attention for environmental sounds in particular. For example,
  the series of challenges on the Detection and Classification of Acoustic
  Scenes and Events (DCASE) \cite{DCASE2013, DCASE2015, DCASE2016, DCASE2017}
  has seen a rapid increase in participation since its first campaign in 2013.
  The number of applications that this area encompasses is extensive, and
  includes query-based sound retrieval \cite{CASSE-retrieval}, smart homes
  \cite{CASSE-smart-homes}, smart cities \cite{CASSE-smart-cities}, and
  bioacoustic scene analysis \cite{CASSE-bioacoustical}.

  Compared to speech and music recognition, the general characteristics of
  environmental sounds are much broader, which means it is difficult to apply
  domain-specific knowledge. Thus, it is important that the method used is able
  to perform well despite little \textit{a priori} knowledge. Supervised deep
  learning methods have largely satisfied this requirement, producing
  state-of-the-art results consistently in this task \cite{crnn-cakir,
  cnn-valenti, attention-xu-kong, gated-crnn-xu}. On the other hand, problems
  such as overfitting have not been completely eliminated, and this is
  especially severe for smaller datasets. To overcome this, we propose a neural
  network architecture based on grouping activation units into
  \textit{capsules} and using a procedure called \textit{routing} during
  inference.

  The notion of a capsule was first introduced in \cite{capsnet-hinton} and
  very recently revisited in \cite{capsnet-sabour} with the addition of a
  routing mechanism. Simply put, a capsule represents a set of properties for a
  particular entity. The authors of \cite{capsnet-sabour} found that routing
  with capsules performed better than the state-of-the-art for digit
  recognition using the MNIST dataset \cite{mnist-lecun}. The motivation for
  capsule routing is that it implicitly learns global coherence by enforcing
  part-whole relationships to be learned. For instance, a person's eye (the
  part) should be positioned sensibly relative to their face (the whole). In
  this case, we would like to associate a capsule representing the eye's
  position to a capsule representing a matching position for the face. If such
  an association cannot be made, it is less likely that a face has been
  identified.

  As a result of this property, capsules overcome shortcomings of current
  solutions such as convolutional networks \cite{mnist-lecun}, which can only
  provide local translation invariance (via max-pooling, typically). In theory,
  routing can introduce invariances for any property captured by a capsule
  \cite{capsnet-sabour}.
 
  It is hypothesized that capsule routing will perform well for SED. One of the
  reasons is contemporary in that current datasets are relatively small, which
  means training is prone to overfitting. To compare with image recognition,
  ImageNet \cite{imagenet-deng} has more than 14 million training samples,
  while most environmental sound datasets have thousands. Indeed, we
  demonstrate this issue in Section \ref{section:results} for a number of
  architectures. By utilizing capsules, we show that overfitting can be
  mitigated.

  A more intrinsic rationale is that capsule routing can be considered as an
  attention mechanism. The idea of attention is to focus on the most salient
  parts of an input via weighting. It has been very successful in numerous
  applications, including machine translation \cite{attention-bahdanau}, image
  captioning \cite{attention-xu}, and, notably, sound event detection
  \cite{attention-xu-kong, gated-crnn-xu}. Attention is particularly useful for
  SED when training data is \textit{weakly labeled}; ground truths for the
  onset and offset times are not available, so the learning algorithm must
  localize sound events without supervision. Routing implements attention by
  weighting the association between lower- and higher-level capsules.

  In this paper, we focus on weakly-labeled event detection. It presents a
  challenge that is relevant to many applications, because collecting labeled
  data is often prohibitively costly. Nevertheless, we believe the main
  contributions of this paper easily apply to the strongly-labeled scenario
  too.

  \section{Capsule Routing}
  \label{section:capsules}
  In general, a neural network is a function \(f: \vec{x} \to \vec{y}\) that is
  composed of several lower-level functions \(f_l: \vec{u} \to \vec{v}\), such
  that \(f = f_L \circ \ldots \circ f_1\). Each lower-level function
  corresponds to a layer in the neural network, and is typically an affine
  transformation followed by a non-linearity, i.e.
  \begin{equation}
    \label{eq:affine}
    \vec{s} = \vec{W}\vec{u} + \vec{b},
  \end{equation}
  \begin{equation}
    \vec{v} = g(\vec{s}),
  \end{equation}
  where \(\vec{W}\), \(\vec{b}\) are learned parameters and \(g(\cdot)\) is a
  differentiable, non-linear function such as the rectifier (ReLU)
  \cite{relu-nair}.

  A capsule network applies the same transformations, but also introduces a
  routing mechanism that affects the learning dynamics. To derive this, we
  rewrite \eqref{eq:affine} as
  \begin{equation}
  \label{eq:affine2}
    \vec{s} =
    \begin{bmatrix}
      \vec{W}_{11}\vec{u}_1 + \ldots + \vec{W}_{1M}\vec{u}_M\\
      \vdots\\
      \vec{W}_{N1}\vec{u}_1 + \ldots + \vec{W}_{NM}\vec{u}_M\\
    \end{bmatrix}.
  \end{equation}
  In \eqref{eq:affine2}, \(\vec{s}\) has been partitioned into \(N\) groups, or
  \textit{capsules}, so that each row in the column vector corresponds to an
  output capsule. Similarly, \(\vec{u}\) has been partitioned into \(M\)
  capsules, where \(\vec{u}_i\) denotes input capsule \(i\), and \(\vec{W}\)
  has been partitioned into submatrices. The bias term, \(\vec{b}\), has been
  omitted for simplicity.

  We now introduce \textit{coupling coefficients}, \(\alpha_{ij}\), so that
  \begin{equation}
  \label{eq:affine3}
    \vec{s} =
    \begin{bmatrix}
      \alpha_{11}\vec{W}_{11}\vec{u}_1 + \ldots
          + \alpha_{M1}\vec{W}_{1M}\vec{u}_M\\
      \vdots\\
      \alpha_{1N}\vec{W}_{N1}\vec{u}_1 + \ldots
          + \alpha_{MN}\vec{W}_{NM}\vec{u}_M\\
    \end{bmatrix}.
  \end{equation}
  Fixing these coefficients to \(\alpha_{ij}=1\) gives \eqref{eq:affine2} and
  hence \eqref{eq:affine}. Instead, we would like these coefficients to
  represent the amount of agreement between an input capsule and an output
  capsule. A capsule encompasses a set of properties, so if the properties of
  capsule \(i\) agree with the properties of capsule \(j\) in the layer above,
  \(\alpha_{ij}\) should be relatively high.
  
\begin{figure}[!t]
    \centering
    \includegraphics[width=0.48\textwidth]{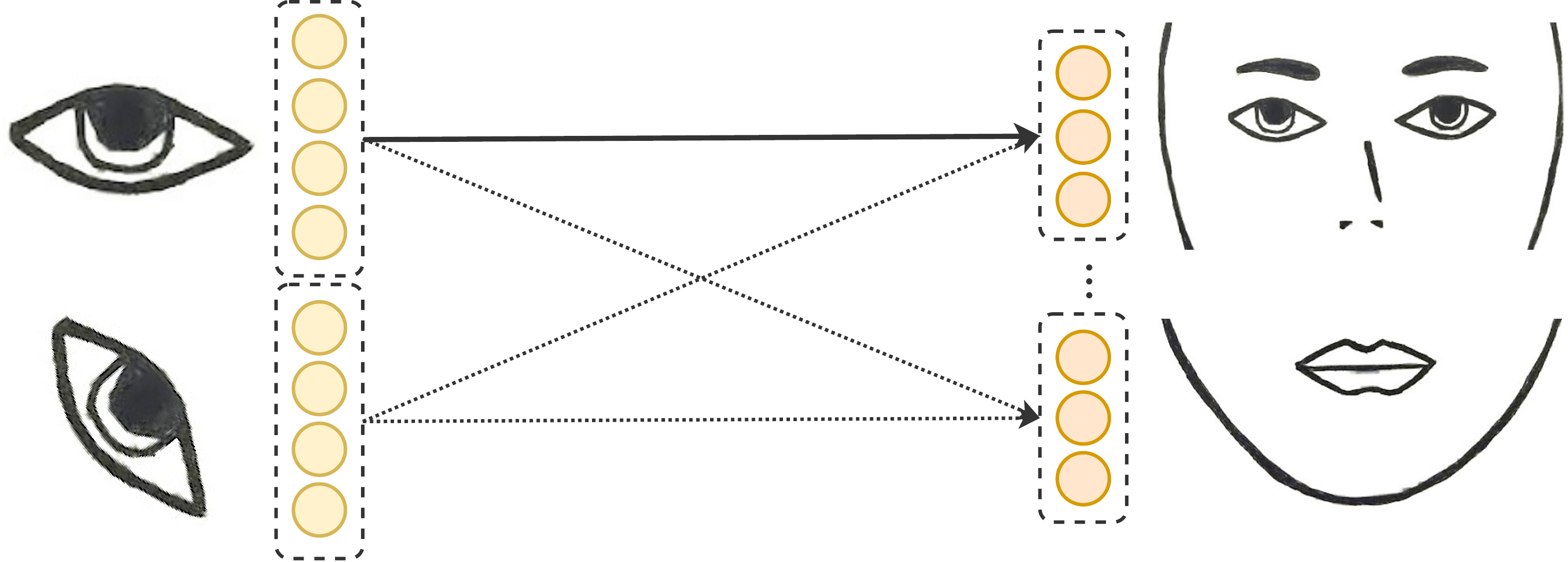}
    \caption{A contrived illustration of the capsule routing concept.
             Activation units are shown as circles and capsules are shown as
             the dashed lines around them. The figure beside each capsule is
             the entity the capsule represents. The capsules in layer \(l\)
             are shown to the left while the capsules in layer \((l+1)\) are
             shown to the right. We see that the correctly-oriented eye
             associates well with the upper face, and is indicated by the
             thick arrow.}
    \label{fig:capsules}
  \end{figure}

  These coefficients are not learned parameters; rather, their values are
  determined using an inference-time procedure called \textit{routing}. The
  idea is based on assigning parts to wholes. Higher-level capsules should
  subsume capsules in the layer below in terms of the entity they identify.
  Routing attempts to find these associations using its notion of agreement,
  which causes the capsules to learn features that enable such a mechanism to
  result in correct predictions. Therefore, global coherencies can be learned
  implicitly, as exemplified in Fig. \ref{fig:capsules}.

  \subsection{Dynamic Routing}
  \label{section:routing}
  Until now, we have given an abstract description of routing. In this section,
  we describe the method used in \cite{capsnet-sabour} to compute the coupling
  coefficients. Noting that a capsule is a vector of activation units, we can
  consider the direction of a capsule as representing its properties. In
  addition, the magnitude of a capsule can be used to indicate how likely it is
  to represent an entity of interest. To ensure that the magnitude is a
  probability, a squashing function is used, and is given by
  \begin{equation}
    \label{eq:squash}
    \vec{v}_j = \frac{\norm{\vec{s}_j}}{1+\norm{\vec{s}_j}^2}
                \frac{\vec{s}_j}{\norm{\vec{s}_j}}.
  \end{equation}

  The method used to compute the coupling coefficients is listed in Fig.
  \ref{alg:routing}. It is a procedure that iteratively applies the softmax
  function to log prior probabilities, \(\beta_{ij}\). These logits are
  initially set to \(\beta_{ij}=0\) to compute \(\vec{v}_j\) and then updated
  based on an agreement computation \(a_{ij} = \vec{v}_j \cdot
  \vec{\hat{u}}_{j|i}\), where \(\vec{\hat{u}}_{j|i} = \vec{W}_{ji}\vec{u}_i\).
  The agreement value is a measure of how similar the directions of capsules
  \(i\) and \(j\) are. The use of the softmax function ensures that \(\sum_j
  \alpha_{ij} = 1\). Thus, \(\alpha_{ij}\) can be seen as the probability that
  the entity represented by capsule \(i\) is a part of the entity represented
  by capsule \(j\) as opposed to any other capsule in the layer above.

  \begin{figure}[!t]
    \vspace{\abovetopsep}
    \hrule height \heavyrulewidth
    \vspace{\belowrulesep}
    \begin{algorithmic}[1]
      \Require Prediction vectors \(\vec{\hat{u}}_{j|i}\),
               layer \(l\), max iterations \(r\)
      \Ensure Layer \((l+1)\) capsules \(\vec{v}_j\)
      \State \textbf{Initialization:} \(\beta_{ij} = 0\)
      \For{\(r\) iterations}
        \State \(\bm{\alpha}_i = \text{softmax}(\bm{\beta}_i)\)
        \State \(\vec{s}_j = \sum_i \alpha_{ij}\vec{\hat{u}}_{j|i}\)
        \State \(\vec{v}_j = \text{squash}(\vec{s}_j)\)
            \Comment cf. \eqref{eq:squash}
        \State \(\beta_{ij} = \beta_{ij} + \vec{v}_j \cdot \vec{\hat{u}}_{j|i}\)
      \EndFor
    \end{algorithmic}
    \vspace{\aboverulesep}
    \hrule height \heavyrulewidth
    \vspace{\belowbottomsep}
    \caption{Routing algorithm. Whenever the indices \(i\) and \(j\) are
             encountered, it should be assumed that it is for all
             \(i = 1 \ldots M\) and \(j = 1 \ldots N\), respectively.}
    \label{alg:routing}
  \end{figure}

  \section{Proposed Method for SED}
  \label{section:method}
  We model the SED task as being comprised of a feature extraction stage and a
  detection stage. Feature extraction refers to transforming the time-varying
  audio signal into a feature vector that is appropriate for subsequent
  detection. The detection stage takes the feature vector as input and attempts
  to detect the sound events that occur and provide timestamps for the start
  and end of each event. This latter stage is where we introduce our neural
  network architecture.

  \subsection{Feature Extraction}
  The input feature vectors are extracted by transforming them to produce a
  logarithmic Mel-frequency (logmel) representation, which is essentially a
  short-time Fourier transform followed by a Mel filterbank and a \(\log\)
  nonlinearity. After this, each resulting feature vector is padded to ensure
  that the inputs to the neural network are of the same dimension. Finally, the
  feature vectors are standardized to zero mean and unit variance. The mean and
  variance parameters used to accomplish this are computed from the training
  set.

  The use of a logmel representation, or the closely-related Mel-frequency
  cepstrum coefficients (MFCC), is standard in the literature due to its good
  performance \cite{asc-barchiesi}. Compared to older techniques such as
  Gaussian mixture models, deep learning benefits from the additional
  information that logmel retains over MFCC. For this reason, we have chosen
  logmel.

  \begin{figure}[!t]
    \centering
    \includegraphics[width=0.45\textwidth]{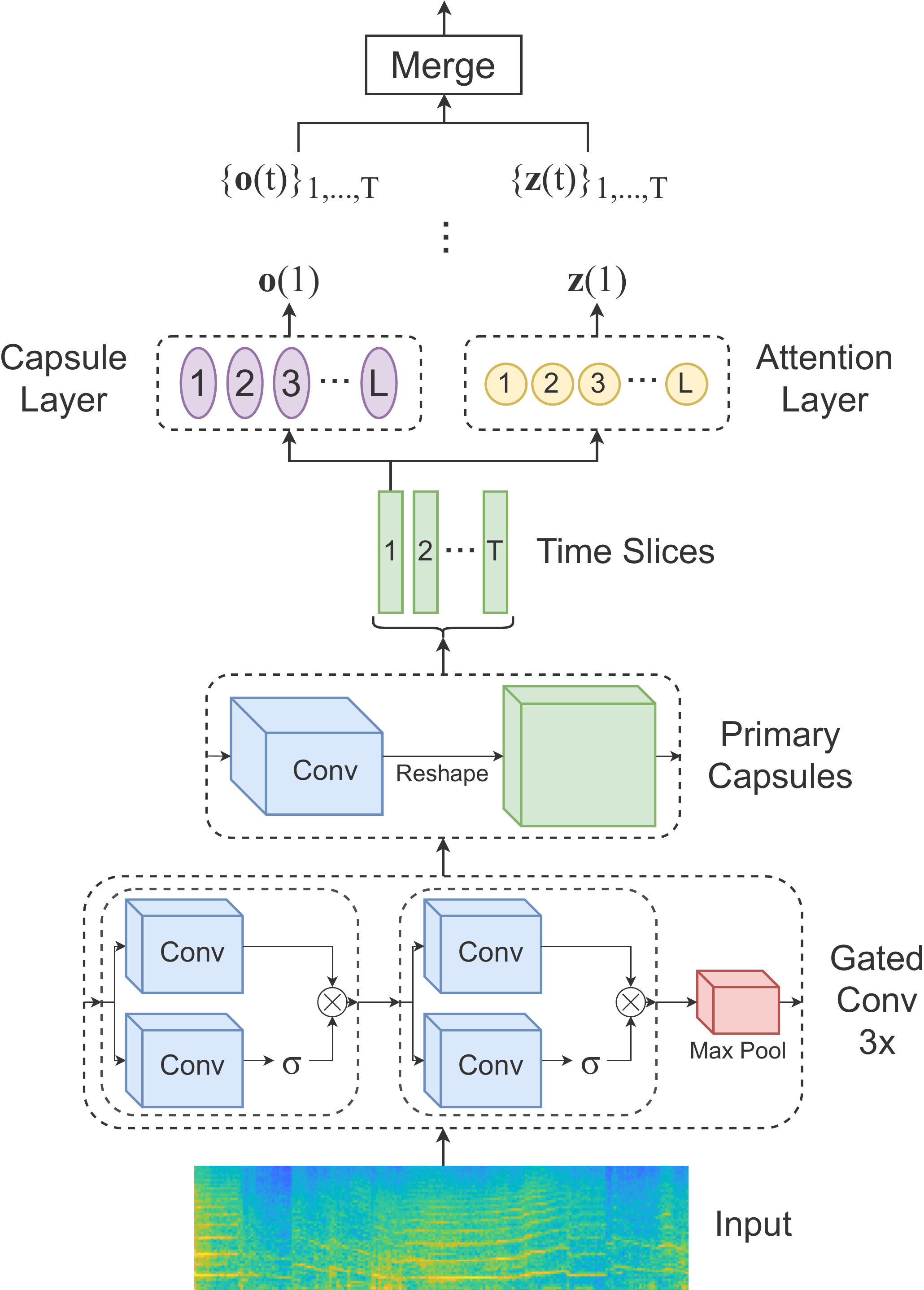}
    \caption{Diagram of the proposed neural network architecture. After the
             primary capsule layer, the output is divided into time slices.
             These slices are transformed by the subsequent layers to give
             \(o(t)\) and \(z(t)\), which are then merged.}
    \label{fig:model}
  \end{figure}

  \subsection{Neural Network Architecture}
  The architecture of the neural network is shown in Fig. \ref{fig:model}. In
  contrast to the ReLU convolutional layer used in \cite{capsnet-sabour}, the
  initial layers of the network are gated convolutions \cite{glu-dauphin,
  gated-crnn-xu}. Experiments showed that a gated nonlinearity improves the
  performance and that having several such layers is beneficial. There are two
  such layers per block and three blocks in total. After each block,
  max-pooling is used to halve the dimensions. The convolutions use 128 filters
  (64 linear, 64 sigmoidal), a kernel width of 3, and a stride of 1.

  Following these initial layers is the primary capsule layer, which is a ReLU
  convolutional layer that has been reshaped into a \(T \times \cdot\footnote{`\(\cdot\)'
  denotes that this dimension can be inferred from the others.} \times U\)
  tensor and squashed using \eqref{eq:squash}. \(T\) is the same temporal
  dimension prior to reshaping and \(U=4\) is the capsule size. In other words,
  each capsule is a \(1 \times 1 \times 4\) slice from the output. The
  convolution uses 64 filters and a kernel width of 3. The stride is set to 1
  for the temporal dimension and 2 for the frequency dimension.

  After this, each \(1 \times \cdot \times 4\) time slice is treated as a
  separate input to the layers that follow. Indeed, the slices are given as
  inputs to two adjacent layers (cf. Fig. \ref{fig:model}): a capsule layer and
  a `temporal attention' (TA) layer. The capsule layer is densely connected
  with \(U=8\) and \(L\) capsules, where \(L\) is the number of classes (sound
  events). Since the previous layer is also a capsule layer, the dynamic
  routing algorithm (Fig. \ref{alg:routing}) is used to compute the output.
  Lastly, the Euclidean length of each output capsule is computed. This gives a
  vector of activations for each time slice \(t\), denoted \(\vec{o}(t) \in
  \mathbb{R}^L\).

  The TA layer is somewhat of a novelty that is not present in the original
  capsule routing paper \cite{capsnet-sabour}. It is used to implement an
  attention mechanism via the saliency of the time slices, and is based on the
  attention scheme described in \cite{attention-xu-kong, gated-crnn-xu}. The
  layer is densely connected with \(L\) units and a sigmoid activation. The
  output is \(\vec{z}(t) \in \mathbb{R}^L\). We can then merge \(\vec{o}(t)\)
  and \(\vec{z}(t)\) across \(t\) so that each prediction, \(y_l\), for class
  \(l\), is given by
  \begin{equation}
  \label{eq:merge}
  \begin{split}
    y_l &= \frac{\sum_{t=1}^T o_l(t) z_l(t)}{\sum_{t=1}^T z_l(t)}\\
              &= \mathbb{E}_{t \sim q_l(t)}[o_l(t)],
  \end{split}
  \end{equation}
  where \(q_l(t) = \text{softmax}(\log \vec{Z}_l)\) and \(\vec{Z}_l \in
  \mathbb{R}^T\) is the collation of \(\{z_l(t)\}_{t=1 \ldots T}\). As such,
  \(y_l\) can be considered as the expected length of the capsule with respect
  to the probability distribution derived from the TA layer. Since \(q_l(t)\)
  is normalized across \(t\), there is an implicit assumption that the sound
  event is present in a single time slice only. Although this is restrictive,
  we justify this choice as a practical compromise, since including the TA
  layer led to better performance in our experiments. In any case, it is
  important to choose an appropriate granularity for the time slices because of
  this.

  Choosing a probability threshold, \(\tau_1\), a sound event \(l\) is present
  if \(y_l>\tau_1\). To calculate onset and offset times, we threshold the
  probabilities of \(o_l(t)\) with another value, \(\tau_2\), and apply a
  morphological closing operation. The purpose of closing is to reduce
  fragmentation and remove noise. The onset and offset times can then be
  determined from the start and end points of the resulting binary regions.

  \section{Experiments}
  \label{section:experiments}
  To evaluate the performance of the proposed method, we used the
  weakly-labeled dataset provided for Task 4 of the DCASE 2017 challenge
  \cite{DCASE2017}. This dataset is comprised of 17 sound event classes, of
  which nine are warning sounds and eight are vehicle sounds. It is divided
  into a training set, a validation set, and an evaluation set, where the
  former contains 51,172 audio clips. Each clip is up to ten seconds in
  duration, and corresponds to one or more sound events that may overlap.

  For this dataset, two tasks were evaluated: audio tagging and sound event
  detection. The former is for detecting which sound events occur in an audio
  clip, while the latter also requires providing onset and offset times. For
  both tasks, performance was evaluated using micro-averages of precision,
  recall, and F-scores. For SED, a segment-based error rate with a one-second
  time resolution was computed too. We used the \textit{sed\_eval} toolbox
  \cite{sed_eval} for evaluation of the SED task. The reader is referred to
  \cite{sed_eval} for a description of these metrics.

  \subsection{System Setup}
  Prior to extracting the features, we resampled each clip to 16 KHz. The
  logmel features were computed using a 64 ms frame length, 20 ms overlap, and
  64 Mel-frequency bins per frame. For a 10-second clip, this gives a \(240
  \times 64\) feature vector.

  To reduce overfitting, we applied batch normalization \cite{batchnorm-ioffe}
  followed by dropout \cite{dropout-hinton, dropout-srivastava} after each
  gated convolutional layer as well as the primary capsule layer. The dropout
  rate (fraction of units to drop) was set to 0.2 for the gated layers and 0.5
  for the primary capsule layer. For capsule routing, the number of iterations
  was set to \(r=3\) following \cite{capsnet-sabour}.

  To train the network, we used binary cross-entropy as the loss function and
  Adam \cite{adam-kingma} as the gradient descent algorithm. The gradient was
  computed using mini-batch sizes of 44. The initial learning rate was set to
  0.001 and decayed by a factor of 0.9 every two epochs. We trained the network
  for 30 epochs, with learned weights being saved per epoch.

  The dataset used in the experiments has a large amount of class imbalance,
  which can lead to bias in the classification. To alleviate this issue, we
  used the data balancing technique suggested in \cite{gated-crnn-xu} to ensure
  that every mini-batch contains a fair number of samples from each class.

  During inference, the five models (epochs) that achieved the highest accuracy
  on the validation set were selected and their predictions were averaged. The
  detection thresholds were set to \(\tau_1=0.3\) and \(\tau_2=0.6\) for our
  system. For SED, the dilation and erosion sizes were set to 10 and 5,
  respectively. As with the other hyperparameters, these values were determined
  based on experiments on the validation set.

  \label{section:results}
  \begin{table}[!t]
    \centering
    \caption{Performance results of audio tagging subtask}
    \label{table:at_results}
    $\begin{tabular}{*{4}{c}}
      \toprule
      Method  & F-score & Precision & Recall \\
      \midrule
      GCCaps  & 58.6\%  & 59.2\%    & 57.9\% \\
      GCNN    & 57.2\%  & 59.0\%    & 57.2\% \\
      GCRNN   & 57.3\%  & 53.6\%    & 59.6\% \\
      EMSI    & 52.6\%  & 69.7\%    & 42.3\% \\
      \bottomrule
    \end{tabular}$
  \end{table}

  \begin{table}[!t]
    \centering
    \caption{Performance results of sound event detection subtask}
    \label{table:sed_results}
    $\begin{tabular}{*{5}{c}}
      \toprule
      Method  & F-score & Precision & Recall  & Error Rate \\
      \midrule
      GCCaps  & 46.3\%  & 58.3\%    & 38.4\%  & 0.76 \\
      GCNN    & 37.5\%  & 46.6\%    & 31.1\%  & 0.88 \\
      GCRNN   & 43.3\%  & 57.9\%    & 34.8\%  & 0.79 \\
      EMSI    & 55.5\%  & -         & -       & 0.66 \\
      \bottomrule
    \end{tabular}$
  \end{table}

  \begin{figure}[!t]
    \centering
    \subfloat[]{
      \includegraphics[width=0.47\textwidth]{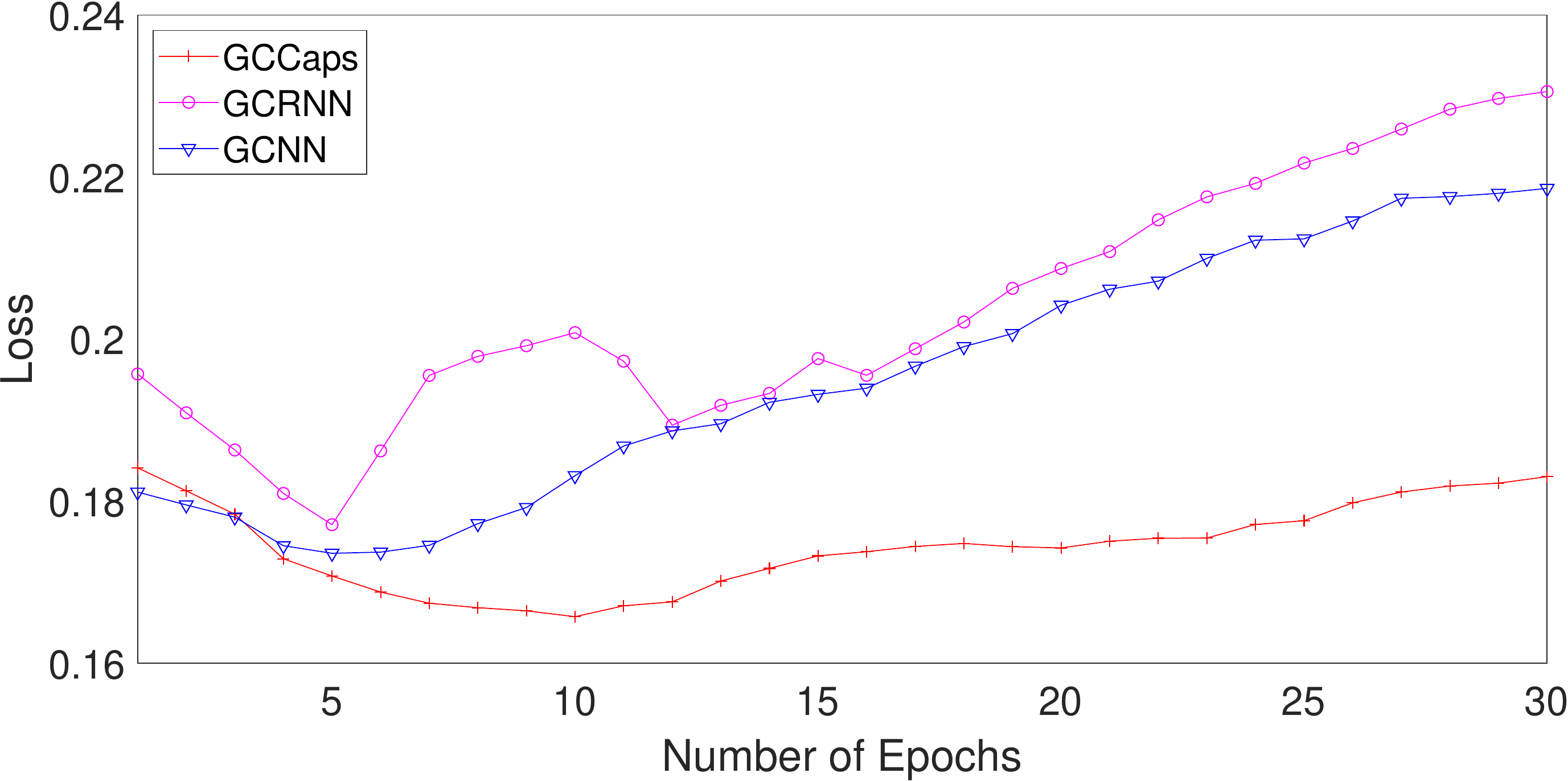}
      \label{fig:loss}}
    \hfil
    \subfloat[]{
      \includegraphics[width=0.47\textwidth]{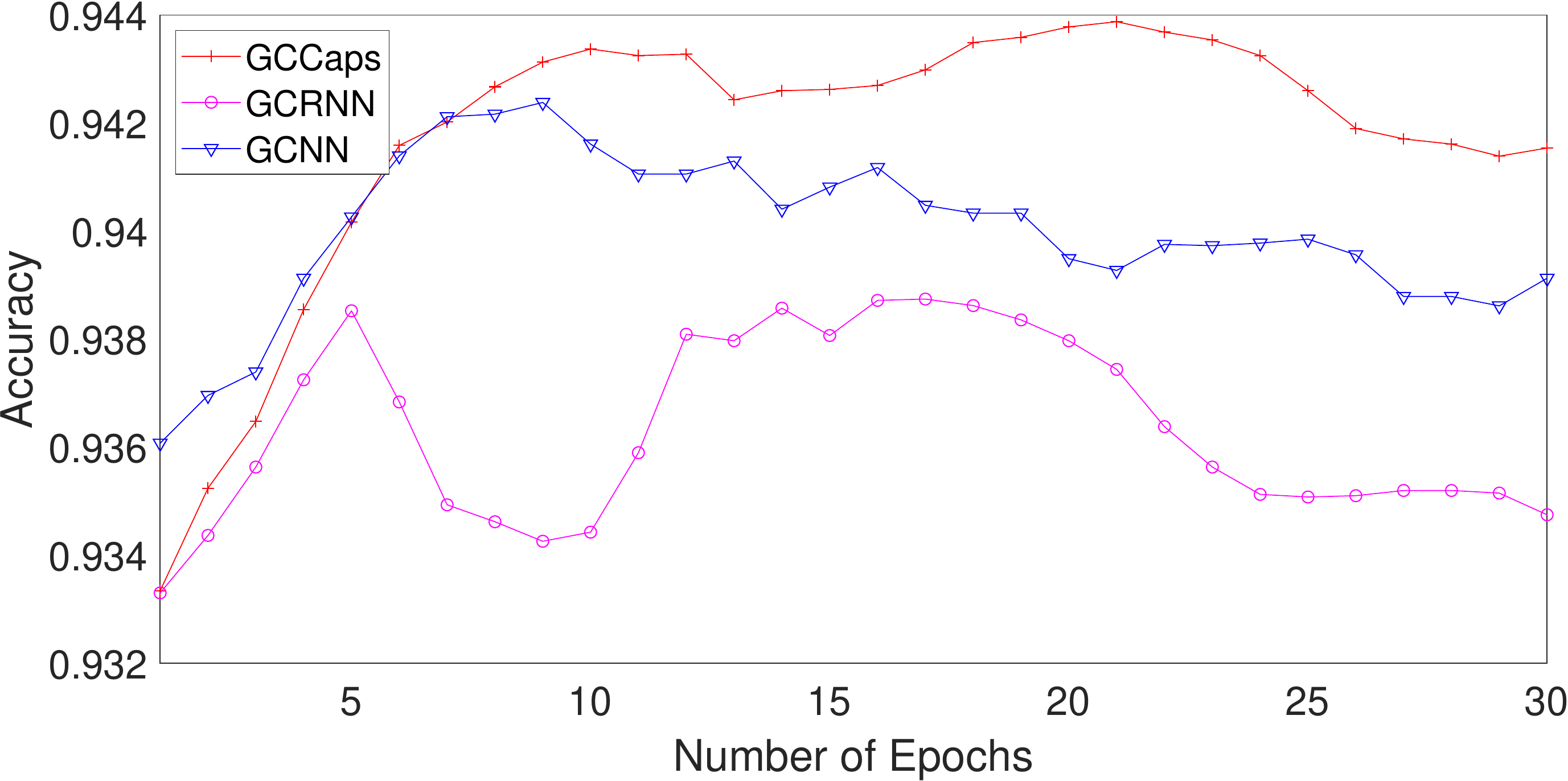}
      \label{fig:accuracy}}
    \caption{Performance as a function of the number of epochs for (a) loss and
             (b) accuracy. The GCCaps model has the highest accuracy and the
             lowest loss, which also does not diverge as severely.}
    \label{fig:epochs}
  \end{figure}

  \subsection{Results}
  In addition to our system (GCCaps)\footnote{Code available online:
  \url{https://github.com/turab95/gccaps}}, we also evaluated the model
  proposed in \cite{gated-crnn-xu} (GCRNN), which won 1\textsuperscript{st}
  place in the audio tagging subtask of Task 4. It is similar to our proposal,
  with the difference being an additional gated convolutional layer and
  recurrent layers \cite{brnn-schuster} as opposed to capsule layers. The same
  model without the recurrent layers (GCNN) is also compared as an ablation
  study for both GCCaps and GCRNN. Moreover, the results for \cite{sed-lee}
  (EMSI) are listed too, albeit much of the setup for that system is not the
  same, so it is not a direct comparison between the architectures. It is
  included because it achieved 1\textsuperscript{st} place in the SED subtask.

  We present our results in Table \ref{table:at_results} and
  \ref{table:sed_results} for audio tagging and sound event detection,
  respectively. For audio tagging, our method performs the best
  overall with an F-score of 58.6\%. EMSI has the highest precision, but its
  recall score is much lower, and, as a result, it has the lowest F-score.
  GCRNN and GCNN perform the same in this subtask.

  For SED, the recurrent layers clearly improve localization for GCRNN, as it
  scores considerably higher compared to GCNN. Meanwhile, GCCaps performs
  marginally better than GCRNN with an F-score of 46.3\% and an error rate of
  0.76. We can deduce that the capsule layers in GCCaps are a good substitute
  for the recurrent layers in GCRNN. EMSI performs the best by a large margin,
  but it should be emphasized that much of the system is different, including
  the use of ensemble techniques to utilize multiple feature vectors, which
  demonstrably \cite{sed-lee} improves its performance significantly.

  To obtain greater insight, we also compared the performance of these models
  (excluding EMSI) on the validation set as a function of the number of epochs.
  As evident in Fig. \ref{fig:epochs}, our proposal achieved the lowest loss
  and highest accuracy, which supports our earlier results. It can be seen in
  Fig. \ref{fig:loss} that all of the models eventually diverge in terms of the
  value of the loss function. In Fig. \ref{fig:accuracy}, it can be seen that
  the accuracy decreases after a number of epochs. These issues are not
  observed with the training set, which suggests that the models are
  overfitting.  However, as shown in the figures, the extent of this problem is
  greatly reduced when using capsule routing. This is reassuring, because it
  indicates that the network can differentiate between fundamental features and
  training-specific features.

  These results demonstrate that a dynamic routing mechanism can improve the
  generalization abilities of a neural network. Although it remains to be seen,
  we are confident that this applies to other datasets too. Investigating
  deeper layers of capsules or different capsule networks, such as
  convolutional capsule networks \cite{capsnet-em, segcaps-lalonde}, is a
  natural direction to take in the future. It is also of interest to explore
  different routing algorithms, such as that proposed in \cite{capsnet-em}.

  \section{Conclusion}
  \label{section:conclusion}
  In this paper, we have proposed a neural network architecture based on
  capsule routing for the detection of sound events. The motivation was that
  capsules can learn to identify global structures in the data that
  alternatives such as convolutional networks cannot. Our system was evaluated
  on a weakly-labeled dataset from Task 4 of the DCASE 2017 challenge. We found
  that the method was considerably less prone to overfitting compared to other
  architectures. For the audio tagging subtask, we achieved a best-in-class
  F-score of 58.6\%, while for the event detection subtask, an F-score of
  46.3\% and an error rate of 0.76. These are promising results, and suggest
  that capsule routing should be further investigated.

  \bibliographystyle{IEEEtran}
  \bibliography{paper}
\end{document}